\documentclass[a4paper, 11pt]{article}
\usepackage{graphicx}
\usepackage{times}
\usepackage{rotating}
\usepackage{natbib}

\begin{document}
\title{Some comments on the paper by Vink et al. 2009 \\ (A\&A, 505, 743 or arXiv:0909.0888)}
\author{Y. Naz\'e (FNRS/ULg), G. Rauw (ULg)}
\maketitle
\abstract{In a recent paper, Vink et al. analyzed some spectropolarimetry data of O-type stars. Here we comment on our disagreement with some points presented in this paper, with the hope of helping to fully grasp the scientific implication of these measurements.}

\section{Introduction}

In a recent draft of a paper on the X-ray emission of an Oe star (Naz\'e et al., see http://arxiv.org/abs/0912.0379v1 ), we made a passing comment that one should exercise caution when referring to the paper of \citet{vin09}. The lead author of that paper wondered about this statement and requested a full explanation of our position. We have a great admiration for J.S. Vink's work in general. However, we do disagree on some key points made in his recent \citet{vin09} paper, and this brief communiqu\'e aims at clarifying our initial comment, with the hope that this will start a healthy scientific discussion leading to a better understanding of the peculiar objects under analysis.

\section{OVz stars}

$\theta^2$\,Ori\,A is mentioned as a spectroscopic binary in the text of p3, yet its spectral type in Table 1 does not provide the associated type of that multiple system (O9.5+B+A) but only its combined type 
(usually only mentioned when the binary nature has not been detected yet) and the polarimetric data are not analyzed taking into account this important fact, as had been done e.g. in X-rays \citep{sch06}.

Among the 17 targets\footnote{or is it 18? Table 1 lists 17 stars for which new data have been obtained, but the text repeatedly mentions the analysis of 18 objects (p2, p11).} of the paper, $\theta^1$\,Ori\,C is certainly THE one where wind asymetries are not put in doubt. Indeed, such asymetries are needed to explain the UV, visible and X-ray behaviour of the star, as recognized towards the end of the paper (p8). However, no depolarization effect is detected in this obvious case, and there is only a short discussion on this lack of detection on p8, where a putative ``inner disk hole" is postulated (line polarimetry could provide constraints on its size, but none is given). MHD simulations explaining the UV-visible-X-ray data have been provided in \citet[][notably Fig. 5 ; see also \citealt{udd08} for further simulations]{gag05}: no ``hole" is seen in the dense equatorial material: close to the star, the rigidly rotating disk is rather dynamic, with continuous infalls/build-ups, but there is always some material around. Admittedly, for faster rotators an ``inner disk hole'' would emerge due to lack of centrifugal support but rotational dynamic effects are minimal for $\theta^1$\,Ori\,C. As such, some explanation is therefore needed on how that ad-hoc guess is compatible with the current evidence. Note that $\theta^1$\,Ori\,C is also a well-known binary \citep[see e.g.][]{pat08}: while this should have no impact on the inner wind structure, well known to be dominated by the magnetic field, this should nevertheless have been mentioned, at least in Table 1.

\section{Of?p stars}

The original reference for the Of?p definition is  \citet{wal72} and a new, more complete definition has been presented in \citet{naz08}. The need for this expanded definition is well explained in Walborn et al. (ApJL, submitted) where Of?p stars are clearly separated from the ``C\,III\,4650-emission-and-nothing-else'' Ofc stars.

On p4, some ``notable changes" of H$\alpha$ are reported/discovered for HD\,148937. This is certainly not new, as these variations (and their 7d period) have been reported previously in \citet{naz08b}, though this is never mentioned.

In Table 2, the old stellar parameters are used for Of?p stars, though results from dedicated new atmosphere modelling are indeed available \cite[see][and references therein]{naz08}. 

On p9, it is mentioned that ``only three O-type stars" exist with a magnetic field detection. Before mid-2009, a few others had been reported \citep{hub08,pet09}, and additional ones have been found since (see e.g. \citealt{gru09}).

On p11, the sentence ``lead us to speculate that it is the presence of a magnetic field that is the underlying reason for this peculiar spectral classification" is certainly no new speculation by the authors, since it was discussed in length in all the recent papers dealing with Of?p stars \citep[see e.g.][quoted in Vink et al.]{don06}.

\section{Oe stars}

Concerning Oe stars, it is stated in Sect. 3.3 that ``double-peaked profiles can for example also be produced in an expanding shell that is spherically symmetric and would, irrespective of the inclination of the stellar rotation axis, not produce any linear polarization". While this is true in principle, such an expanding shell would be expected to produce emission lines that are strongly variable on the expansion time scale (which should be of the order of weeks at most, taking into account the peaks' velocities). Indeed, the decrease of the density of an expanding wind would lead to dramatic variations of the equivalent widths of the emission lines. However, this has not been observed. For example, the variability of two Oe stars (HD45314 and HD60848) has been studied by \citet{rau07}. Whilst the equivalent widths of H$\alpha$ lines in these objects change, the time scales of these variations are much longer and the amplitudes far less extreme than expected from the above considerations. It is therefore unclear how the proposed ``shell" scenario agrees with the observational evidence. 

Generally speaking, the discussion about Oe stars in Sect. 3.3 and 4.3 does not take into account the information that comes from the temporal spectroscopic variability of these objects \citep[e.g.][]{rau07}. This variability is very similar to that observed in Be stars where it is frequently attributed to density waves in the circumstellar disk \citep[e.g.][]{han95,por03}. This variability needs to be taken into account for a discussion of the presence or absence of disks around Oe stars. 

\section{Onfp stars}

The ``sample of Onfp supergiants" studied by the authors is not adequate to make a statistical analysis (such as quoted in the abstract ``3/4 group IV Onfp stars show evidence for complex polarization effects likely related to rapid rotation", see also Table 1 \& 2, Sect. 3.4, 4.5 and 5) and/or to draw any conclusion about genuine Onfp stars. In fact, while the 4 ``Onfp stars" considered by the authors were indeed initially classified in this category (using photographic spectra in the early 1970s), this classification is clearly outdated for two of them (HD152248 and Cyg OB2 \#5) and no longer supported by the present-day knowledge of these objects. 

Cyg OB2 \#5 is an eclipsing interacting binary system that has been intensively studied before \citep{boh76,rau99,lin09}. These studies revealed in fact that the morphology of the He\,{\sc ii}\,4686 line, that might have motivated an Onfp classification in older studies based on photographic plates, is highly variable with binary phase (Figs. 3 and 13 in \citealt{rau99}) and reflects actually the complex flows of matter in this binary system. The same remark applies to the H$\alpha$ line (Fig.3 in \citealt{lin09} and Figs. 1 \& 2 of \citealt{vre85}). Under these circumstances, it is obvious that this star can by no means provide any information about genuine Onfp stars.

The situation is very much similar for the eclipsing binary HD152248. \citet{san01} clearly showed that the He\,{\sc ii}\,4686 and H$\alpha$ lines are variable with orbital phase (Fig. 7 in \citealt{san01}) and that the morphology of the profile is to a large extent determined by the orbital motion of the stars and the wind-wind interaction in this system. From the analyses of the light curve of HD152248 published by \citet{may08}, it is also quite obvious that none of the components of this system is a supergiant. With a true spectral type of O7.5III(f)+O7III(f)  \citep{san01}, no star in HD152248 appears to be a supergiant or nfp star: it is therefore obvious that this star cannot provide any information about genuine Onfp supergiants.

In the ``group IV sample", the Onfp classification hence only applies to $\lambda$\,Cep and $\zeta$\,Pup. As for Oe stars, the spectroscopic variability of these objects is well known (e.g. the variability of the He\,{\sc ii}\,4686 line discussed by \citealt{mof81} and \citealt{kap99}) but never discussed, though this variability is usually attributed to large scale structures in the winds of these stars and is therefore clearly relevant here. Note also that variability studies exist in the literature for the full sample of genuine galactic Onfp (or Oef in the terminology of \citealt{con74}) stars: $\zeta$\,Pup, $\lambda$\,Cep, BD+60$^{\circ}$2522, HD14434, HD14442 and HD192281. \\

{\small {\it Acknowledgments: } We thank A. ud-Doula for his help in writing these comments.}

\end{document}